Dear Sir,
here enclosed please find the manuscript:

ELASTIC CONSTANT DISHOMOGENEITY .... IN DISORDERED SYSTEMS

by M. Montagna et al

that I submit for publication in Physical Review Letters.
5 PstScript figures are being sent as separate files.

We suggest that one of the referees could be chosen between
P. Allen (NY State University at Stony Brook) and J. Feldman
(Naval Research Laboratory, Washington DC).

With best regards

G. Viliani
Dipartimento di Fisica 
Universita' di Trento
38050 Povo, Trento
Italy

Phone: +39 0461 881555
Fax:   +39 0461 881696

\documentstyle[twocolumn,prl,aps]{revtex}
\begin{document}
\draft
\twocolumn[
\hsize\textwidth\columnwidth\hsize\csname @twocolumnfalse\endcsname
\title
      {
	Elastic constant dishomogeneity and $Q^2$ dependence of the
	broadening \\
	of the dynamical structure factor in disordered systems
      }
\author{ 
	 M.~Montagna$^1$,          
	 G.~Ruocco$^2$,  
	 G.~Viliani$^1$, 
	 R.~Di Leonardo$^2$, 
	 R.~Dusi$^1$,  
	 G.~Monaco$^2$, 
	 M.~Sampoli$^3$,
	 T.~Scopigno$^1$. 
	}
\address{
	 $^1$
	 Dipartimento di Fisica and INFM, Universit\'a di Trento,
	 I-38100, Povo, Trento, Italy.\\
	 $^2$
	 Dipartimento di Fisica and INFM, Universit\`a di L'Aquila,
	 I -67100, L'Aquila, Italy. \\
	 $^3$
	 Dipartimento di Fisica and INFM, Universit\`a di Firenze,
	 I -50139, Firenze, Italy. \\
	}

\date{\today}
\maketitle
\begin{abstract}
We propose an explanation for the quadratic dependence on the momentum $Q$, of 
the broadening of the
acoustic excitation peak recently found in the study of the dynamic structure 
factor of many real and simulated glasses. We ascribe the observed $Q^2$
law to
the spatial fluctuations of the local wavelength of the collective vibrational 
modes, in turn produced by the dishomegeneity of the inter-particle elastic 
constants. This explanation is analitically shown to hold for 1-dimensional 
disordered chains and satisfatorily numerically tested in both 1 and 3
dimensions.
\end{abstract}


]

The study of the nature of collective atomic excitations in disordered 
solids at wavelengths $\lambda$ approaching the interparticle separation
$a$, $a/\lambda \approx 0.1-0.5$, has received renewed interest in the 
last few years thanks to new experimental tools and to improved numerical 
techniques. On the experimental side, the dynamics of the collective 
excitations is often investigated via the dynamic structure factor
$S(Q,\omega)$, i.~e. the space-time Fourier transform of the particle
density correlation function. The study of $S(Q,\omega)$ in the region 
of mesoscopic exchanged momentum ($Q=1-10$ nm$^{-1}$) has become recently 
possible in disordered systems thanks to the development of the Inelastic 
X-ray Scattering (IXS) technique \cite{IXS}, and many glasses \cite{Glass} 
and liquids \cite{Liq} have been studied with this technique. Although 
specific {\it quantitative} differences exist among different systems, 
all the investigated glasses show some common features that can be summarized 
as follows: {\it i)} there exist propagating acoustic-like excitations up to 
$Q$$a \approx 3$;
{\it ii)} the slope of the $Q-\omega$ dispersion relation
in the $Q\rightarrow 0$ limit extrapolates to the macroscopic sound
velocity; {\it iii)} the broadening of the excitation peaks in  
$S(Q,\omega)$ follows a $Q^2$ law.
These general features of $S(Q,\omega)$ have been confirmed by numerical
calculations in different glasses, using both standard Molecular Dynamics 
(MD) simulations \cite{MDglass}, and the Normal Mode Analysis (NMA) in the 
harmonic approximation \cite{NMA}.

The $Q^2$ dependence of the excitations broadening has not yet received 
a theoretical explanation. Such dependence is the same as predicted by 
hydrodynamics, but this coincidence is only accidental. Indeed, in the 
experiments the broadening is found to be temperature independent and 
the $Q^2$ law is numerically found also in {\it harmonic} glass models; 
these results indicate that the  origin of this behavior should be found 
in {\it structural} rather than {\it dynamical} properties, i.~e. it
should be associated to the atomic disorder in the glass and not to 
dissipative phenomena like anharmonicity or relaxation processes. 
For this reasons, we concentrate on harmonic systems, leaving apart 
all the difficulties due to dynamical processes like, for example, 
anharmonicity. 

In this Letter we suggest that the observed broadening 
is due to the spatial fluctuations of the elastic constants.
Our explanation is based on analytical and numerical results for the simple
case of a 1-dimensional (1-d) harmonic disordered system (linear chain of
like mass joined by random springs), and then it is generalized to 3
dimensions.

The dynamics of 1-d disordered lattices has been thoroughly investigated in
the past,and a recent overview on this subject can be found in \cite{CPV}. 
In principle, 1-d systems often show very peculiar characteristics; more
specifically it  is well known \cite{1dLoc} that, at variance with d$>$1 
cases, in a (infinite) disordered linear chain all vibrational modes are 
localized and localization by itself contributes to the broadening of 
the inelastic peaks of $S(Q,\omega)$. However, as we will show,
this effect can be disentangled.
Our model consists of an linear chain of $N$ particles ($i=1..N$) of
mass $M$ placed a distance $a$ apart from each other ($x_i=ia$), and joined 
by next-neighbour springs ($K_i$) randomly chosen from a flat 
distribution (${\cal P}(K)$) with extrema $K \pm \Delta K$ (mean 
$\mu_k=K$ and standard deviation $\sigma_k=\Delta K/\sqrt{3}$).
The dynamics of the system can be expressed in terms of its 
eigenvalues ($\omega_p$) and eigenvectors ($e_p(i)$ ), $p=1...N$ being the
mode label. The classical dynamic structure factor is expressed as:
\begin{equation}
\label{s}
S(Q,\omega)=\frac{K_BT}{M} \frac{Q^2}{\omega^2}E(Q,\omega)
\end{equation}
with
\begin{eqnarray}
\label{e}
E(Q,\omega)=\sum_p \vert \tilde e_p(Q) \vert^2 \delta(\omega-\omega_p), \\
\nonumber
\tilde e_p(Q)=N^{-1/2} \sum_i \exp(iQx_i) e_p(i).
\end{eqnarray}

The eigenvectors and eigenvalues are the solutions of the secolar problem,
$\sum_j D_{ij} e_p(j)=\omega^2_p e_p(i)$ where $D_{ij}$ is the 
dynamical matrix. The basic properties of $S(Q,\omega)$ are cointained 
in $E(Q,\omega)$, the squared space Fourier transform of the eigenvectors
 of modes
with frequency $\omega_p$ "close" to $\omega$. A typical eigenvector obtained 
from the diagonalization of the dynamical matrix is reported in Fig.~1. 
As is well known (see for example \cite{ishii}),
there exists a rather well defined wavelength but, as compared with
the sinus function expected in the case of an ordered system, three main 
differences can be noted: {\it i)} the peak height is not constant, i.~e. 
there exists an envelope which is localized in space (see inset a); 
{\it ii)} the wavelength $\lambda$ (i.~e. the distance between two next 
nearest nodes) is not constant, but is rather a space fluctuating quantity;
its statistical distribution is reported in the inset b) of Fig.~1; finally, 
{\it iii)} by analysing in detail the eigenvector between two successive
 nodes
(see insets c) and d), some deviations from the simple sinus law can be
evidenced.
In principle, all these three characteristics could contribute to broade the 
inelastic peaks in $S(Q,\omega)$. In the following we will show that the 
broadening is mainly due to fluctuations of the local wavelength and (in 
1-d systems) to localization, while {\it iii)} has minor consequences, and 
its signature can only be found in the low-$\omega$ tail of $S(Q,\omega)$.

Let us first focus on {\it ii)}, i.~e. on the effect of spatial wavelength 
fluctuations. 
To show how these fluctuations by themselves can produce the $Q^2$-broadening,
we
build up a model eigenmode which possess only characteristic {\it ii)}, 
that is a sequence of alternating positive and negative semiperiods of a
sinus function of slightly different wavelength. If $h_m$ ($m=1,2...$)
are the positions of the nodes, we define the local wavelengths 
$\lambda_m=h_m-h_{m-2} $ and compute the Fourier transform of the model 
eigenvectors, i.e. $ \tilde e(Q) \propto \sum_m  \int_{h_{m-2}}^{h_m} dx \; 
\exp(iQx) \; \; \sin(2 \pi/\lambda_m(x-h_{m-2})) $. This is then averaged over 
different realizations of the disordered chain, assuming a gaussian
distribution
of wavelengths centered at $\lambda$ and with variance $\sigma^2_\lambda$. 
The calculation of $E(Q)=N^{-1} \langle |\tilde e(Q)|^2 \rangle $ is 
straightforward althought long; it produces an $E(Q)$ that near the 
peak, located at $Q=2\pi /\lambda$, has a quasi-lorentzian shape with half 
width at half maximum (HWHM) $\Gamma^{^{(Q)}}_{_{F}}$ given by: 
\begin{equation}
\label{dq1}
\frac{\Gamma^{^{(Q)}}_{_{F}}}{Q} = {\pi}\frac{\sigma^2_\lambda}{\lambda^2}
\end{equation}
The successive step is to find a relationship between 
$\sigma_\lambda$ and the characteristics of disorder. In each segment of 
length $\lambda_m$, containing $n_m$ springs, we define a "local" sound 
velocity, $v^{loc}_m = a \sqrt{K_m^{eff} / M}$, where $K_m^{eff}$ is the
effective elastic constant obtained by averaging the individual spring
constants $K_i$ that are found between $h_{m-2}$ and $h_{m}$: 
$(K_m^{eff})^{-1}= \sum_{i \in m} {P_i}K_i^{-1}/\sum_{i \in m}P_i$. 
Here the $P_i$'s are weights that take into account the fact that springs
near the nodes are more effective (highly stretched) than those near the 
antinodes (not stretched) in determining $K_m^{eff}$.
The validity of the concept of "local velocity" can be tested numerically.
In fact, coming back to the eigenvectors obtained by the solution of the
secular
problem for the disordered linear chain, the actual wavelength
of the $m$-th half-period, $\lambda_m$, can be compared with
 that obtained from the local 
sound velocity, i.~e. from the relation $\lambda^{loc}_m = 2\pi
v^{loc}_m / \omega$.
As an example, in Fig. 2 we show the correlation between $\lambda$ and 
$\lambda^{loc}$ obtained from the analysis of the eigenvectors of the modes 
at $\omega/\sqrt{K/M} \approx 0.031$ of 1,000 disordered chains with 
$\Delta K / K $=0.3, for different choices of the weights $P_j$.
As can be seen,
the correlations are satisfactory, indicating the validity of the
assumptions
that {\it i)} a local sound velocity exists, and {\it ii)} it is
determined by a
local spring constant averaged over the wavelength. Moreover, we observe
that the
highest correlation is found for weights proportional to the square of the
local strain. From the previous relations it is easily deduced that:
\begin{equation}
\label{hh}
\frac{\sigma_\lambda}{\lambda} = 
	\frac{\sigma_v}{v} =
	\frac{1}{2} \frac{\sigma_{K^{eff}}}{K^{eff}} =
	\frac{1}{2} \sqrt{\frac{\alpha}{n}} \frac{\sigma_k}{\mu_k}
\end{equation}
where $n=\langle n_m \rangle = \lambda/a$ and $\alpha = \langle P^2 \rangle/ 
\langle P \rangle^2$. In the following we will choose $P$ proportional to
the square
of the strain and, therefore, $\alpha=3/2$. By substituting Eq.~\ref{hh}
into \ref{dq1}, we obtain:
\begin{equation}
\label{dq2}
\Gamma^{^{(Q)}}_{_{F}} = a \frac{\sigma_k^2}{\mu_k^2} \frac{\alpha}{8} Q^2
\end{equation}
which reproduces the experimental $Q^2$ behavior. Summing up, we have
shown here
that the spatial elastic constant fluctuations by themselves produce a
broadened $E(Q,\omega)$ which is quasi-lorentzian in shape with HWHM
$\Gamma^{^{(Q)}}_{_{F}} \propto Q^2$.

In order to check the validity of Eq. \ref{dq2} we performed a numerical
calculation of $E(Q,\omega)$. At selected values of $\omega$, and for
different values of $\Delta K$, we calculated the eigenvectors of 50
different realizations of a disordered chain composed of 20,000 atoms using
the Dyson-Schmidt (DS) method \cite{DS}. 
Using Eq.~\ref{e} the functions $E(Q,\omega)$ are then calculted as a
function of $Q$; representative examples are reported in Fig.~3 for the
indicated values of $\omega$ and $\Delta K / K=0.6$. 
%
%
The $E(Q,\omega)$ have been fitted to a lorentzian lineshape. The derived
HWHM $\Gamma$ are reported in Fig.~4 together with our prediction 
(Eq.~\ref{dq2}), which reproduces correctly the $Q^2$ behavior but 
is about a factor 2 too small. This discrepancy is due to our neglecting 
of localization effects which, as mentioned before, are certainly effective
in 1-d systems.  In these systems the eigenvectors 
have an exponential envelope in the tails, whith a decay length 
$L$, and this produces a contribution to the linewidth of $E(Q,\omega)$ 
vs $Q$ given by $\Gamma^{^{(Q)}}_{_{L}} = 1/L$ \cite{CPV}. Since
localization
also gives rise to a lorentzian lineshape, we expect a total broadening
given by
$\Gamma^{^{(Q)}}_{_{Tot}} = \Gamma^{^{(Q)}}_{_{L}} + \Gamma^{^{(Q)}}_{_{F}}$. 
The values of $L$ have been numerically computed by fitting the tails of the 
eigenvectors produced by the DS technique; as expected \cite{CPV} we found 
$L \propto \omega^{-2}$ and, therefore, also in this case 
$\Gamma^{^{(Q)}}_{_{L}} \propto Q^2$. 
The full line in Fig. 4 represents the total HWHM $\Gamma^{^{(Q)}}_{_{Tot}}$: 
it is in quantitative agreement with the widths of $E(Q,\omega)$ determined
numerically.

We stress that the effect of localization is peculiar to 1-d
systems and is not relevant to 3-d disordered systems \cite{NMA}. Therefore,
we expect that in 3-d the wavelength fluctuations induced
by the spatial flutuations in the local elastic constant account for most
of the broadening observed in experiments. Since usually experiments
probe $S(Q,\omega)$ as a function of $\omega$ at fixed $Q$, we  
rewrite Eq. \ref{dq2} as 
\begin{equation}
\label{dom}
\Gamma^{^{(\omega)}}_{_{F}}= v \Gamma^{^{(Q)}}_{_{F}} 
= a\;v \frac{\sigma_k^2}{\mu_k^2} \frac{\alpha}{8} Q^2
\end{equation}
where $v$ is the sound velocity. In order to verify whether this equation,
derived for 1-d, is still valid for 3-d systems we determined numerically 
the broadening of $S(Q,\omega)$ for a simple model glass consisting 
of 32,000 Ar atoms interacting via the 12-6 Lennard-Jones (LJ) potential 
($\sigma=3.405$ $\AA$, $\epsilon/K_B= 125.2 $ K); for further details on  
this simulation see \cite{PMAr}. 
In the inset of Fig.~5 we report the $\Gamma^{^{(\omega)}}$ values measured
from a lorentzian fit to the numerically calculated $S(Q,\omega)$ (which are
also shown in Fig.~5), compared with the prediction of Eq.~\ref{dom}
(full line).
The values of $\sigma_k$, $\mu_k$ and $a$, needed to calculate 
$\Gamma^{^{(\omega)}}_{_{F}}$ from Eq.~\ref{dom}, were determined by averaging
the interatomic distances and the second derivatives of the LJ interatomic 
potential \cite{note}.  The good quantitative agreement between the numerical 
and the predicted values of $\Gamma^{^{(\omega)}}$ gives us confidence
on the applicability of Eq.~\ref{dom} to real glasses, and therefore on
the prevalence of the elastic-constant fluctuation mechanism in 
determining the width of inelastic peaks in the $S(Q,\omega)$. 

We emphasize that the mechanism proposed for the 
origin of the broadening of the Brillouin peaks in $S(Q,\omega)$ is 
completely different from the situation where a mode with well defined
wavelength
exists and its scattering is produced by isolated point defects (Rayleigh 
scattering), which would result in a $Q^4$ law. 

In conclusion, we propose a plausible explanation for the ubiquitous 
$Q^2$ law found for the broadening of the Brillouin peaks in disordered
systems.
According to the present model this broadening is a consequence of the 
spatial fluctuation of the "effective" (i.~e. averaged over one wavelength) 
elastic constant. This fluctuation becomes smaller and smaller as the
wavelength
(i.~e., as the number of involved springs) increases and in particular,
from
basic statistics it is $\propto \lambda^{-1/2}$. This result, together 
with the existence of a "local velocity", leads to a $Q^2$ dependence of the 
Brillouin linewidth in harmonic disordered systems at all $Q$'s.

\begin{center}
{\bf FIGURE CAPTIONS}
\end{center}
{
\footnotesize {
\begin{description}
\setlength{\leftmargin}{0cm}
\item {FIG. 1
Portion of the eigenvector of the mode at $\omega/\sqrt{K/M}=0.571$ of a 
chain with N=1,000 particles and $\Delta K/K=0.6$ reported as a function of 
the particle coordinate. In the insets the three most relevant 
features of
the eigenvectors are emphasized: 
{\it i)} Localization (the envelope of the whole eigenvector
is reported in inset a); {\it ii)} Wavelength fluctuations (the distribution
of $\lambda_m$ calculated from 100 realization of disorder is reported in 
inset b together with the gaussian fit, $\bar\lambda/a$=10.18, 
$\sigma_\lambda/a$=0.827); and {\it iii)} Deviation from sinus curve (a blow
up of the eigenvector, dots, is reported in the inset c) together with the best
local sinus approximation, full line. In the inset d) it is reported the residual 
of the eigenvector after the subtraction of the full line in the inset c).)
}

\item {FIG. 2 
Countour plot of the numerical joint distribution function 
${\cal{P}}(\lambda,\lambda^{loc})$ obtained by the analysis of the 
modes at $\omega/\sqrt{K/M}\approx0.031$. The countour plot has  
been obtained averagin 1,000 realization of disordered chains 
($\Delta K/K=0.3$) with $N=$1,000. The analysis has been performed 
with different choices for the weigth: 
a) $P$=1; b) $P=\vert \nabla e_p(i) \vert$; and 
c) $P=\vert \nabla e_p(i) \vert^2$. The $\cal{P}$-scale is logarithmic and
the $i$-th line indicates the level at $10^{-i/4} {\cal{P}}_{max}$.
}

\item {FIG. 3  
Examples of $E(Q,\omega)$ vs $Q$ at the indicated $\omega$ values 
calculated with the Dyson-Schmidt method \cite{DS} averaged over 50
realizations of disordered linear chains ($\Delta K/K=0.6$) of length
$N=20,000$. Only the peak region is reported, and the full lines are
the best lorentzian fits in this region.
}

\item {FIG. 4  
$\omega$-dependence of the hwhm of the lorentzian fit to the $E(Q,\omega)$
reported in Fig.~3 (full dots) compared with the predictions from elastic
constant fluctuations including (full line) or not including (dashed line)
the localization effects.
}

\item {FIG. 5 
Examples of $S(Q, \omega)$ of simulated glassy Argon at the indicated $Q$ values.
In the inset the hwhm of the peaks of the $S(Q,\omega)$, calculated by
a lorentzian fit in the peak region, (full dots) are reported vs $Q$ together
with the theoretical prediction from Eq.~\ref{dom} (full line).
}
\end{description}
}}


\begin{references}
\bibitem{note}
The quantities $a$, $\mu_k$, and $\sigma_k$, entering Eq.~\ref{dom}, were
determined by the equations:
$$
\langle O \rangle = \left[ \int_o^{r_o} dr \; 4 \pi \; r^2 \; g(r)\; O(r) \right ]
\left[ \int_o^{r_o} dr \; 4\pi \; r^2 \; g(r) \right ]^{-1}
$$
where $O(r)=r$ for $a$, $O(r)=V^{''}_{_{LJ}}(r)$ for $\mu_k$ and 
$O(r)=(V^{''}_{_{LJ}}(r)-\mu_k)^2$ for $\sigma^2_k$. Here $V_{_{LJ}}(r)$
is the LJ potential function,  $g(r)$ is the pair distribution function of 
the system and $r_o$ a cut-off radius that takes into account that in the 
theoretical derivation of Eq.~\ref{dom} we only considered nearest neighbours 
interaction. The results only slightly depend on the choice of $r_o$, 
and we fixed it at half way between the first and second peak in the $g(r)$.
\end{references}
\end{document}